\documentclass{elsarticle}

\usepackage{lineno}
\usepackage{hyperref}
\usepackage{bm}
\usepackage{color}
\usepackage{amsmath}
\usepackage{amssymb}
\usepackage{float}
\usepackage[top=0.85in,left=1in,footskip=0.75in]{geometry}

\journal{Chaos, Solitons \& Fractals}

\modulolinenumbers[3]

\biboptions{sort,compress}
\makeatletter
\providecommand{\doi}[1]{%
  \begingroup
    \let\bibinfo\@secondoftwo
    \urlstyle{rm}%
    \href{http://dx.doi.org/#1}{%
      doi:\discretionary{}{}{}%
      \nolinkurl{#1}%
    }%
  \endgroup
}
\makeatother
\let\oldequation\equation
\let\oldendequation\endequation

\renewenvironment{equation}
  {\linenomathNonumbers\oldequation}
  {\oldendequation\endlinenomath}
  
\begin{document}

\begin{frontmatter}

\title{Integrating theory and experiments to link local mechanisms and ecosystem-level consequences of vegetation patterns in drylands}

\author[add1]{Ricardo Martinez-Garcia}
\ead{ricardom@ictp-saifr.org}
\author[add2,add3]{Ciro Cabal}
\author[add4,add5,add6]{Justin M. Calabrese}
\author[add7]{Emilio Hern\'andez-Garc\'ia}
\author[add2]{Corina E. Tarnita}
\author[add7]{Crist\'obal L\'opez}
\author[add8]{Juan A. Bonachela}

\address[add1]{ICTP South American Institute for Fundamental Research \& Instituto de F\'isica Te\'orica - Universidade Estadual Paulista, S\~ao Paulo, SP, Brazil}
\address[add2]{Department of Ecology and Evolutionary Biology, Princeton University, Princeton, NJ, USA.}
\address[add3]{Department of Biogrography and Global Change, National Museum of Natural Sciences, MNCN, CSIC, Madrid, Spain.}
\address[add4]{Center for Advanced Systems Understanding (CASUS), Görlitz, Germany}
\address[add5]{Helmholtz-Zentrum Dresden Rossendorf (HZDR), Dresden, Germany}
\address[add6]{Department of Ecological Modelling, Helmholtz Centre for Environmental Research – UFZ, Leipzig, Germany}
\address[add7]{IFISC, Instituto de F\'isica Interdisciplinar y Sistemas Complejos (CSIC-UIB), Palma de Mallorca, Spain}
\address[add8]{Department of Ecology, Evolution, and Natural Resources, Rutgers University, New Brunswick, NJ, USA}

\begin{abstract}
Self-organized spatial patterns of vegetation are frequent in drylands and, because pattern shape correlates with water availability, they have been suggested as important indicators of ecosystem health. However, the mechanisms underlying pattern emergence remain unclear. Some theories hypothesize that patterns could result from a water-mediated scale-dependent feedback (SDF) whereby interactions favoring plant growth dominate at short distances and growth-inhibitory interactions dominate in the long range. However, we know little about how the presence of a focal plant affects the fitness of its neighbors as a function of the inter-individual distance, which is expected to be highly ecosystem-dependent. This lack of empirical knowledge and system dependency challenge the relevance of SDF as a unifying theory for vegetation pattern formation. Assuming that plant interactions are always inhibitory and only their intensity is scale-dependent, alternative theories also recover the typical vegetation patterns observed in nature. Importantly, although these alternative hypotheses lead to visually indistinguishable patterns, they predict contrasting desertification dynamics, which questions the potential use of vegetation patterns as ecosystem-state indicators. To help resolve this issue, we first review existing theories for vegetation self-organization and their conflicting predictions about desertification dynamics. Second, we discuss potential empirical tests via manipulative experiments to identify pattern-forming mechanisms and link them to specific desertification dynamics. A comprehensive view of models, the mechanisms they intend to capture, and experiments to test them in the field will help to better understand both how patterns emerge and improve predictions on the fate of the ecosystems where they form.
\end{abstract}


\end{frontmatter}

\linenumbers

\section{Introduction}
From microbial colonies to ecosystems extending over continental scales, complex biological systems often feature self-organized patterns, which are regular structures that cover large portions of the system and emerge from nonlinear interactions among its components \citep{Meinhardt_1982,Camazine_2003,Sole_2006,Pringle_2017,MartinezGarcia2022}. Importantly, because harsh environmental conditions provide a context in which self-organization becomes important for survival, emergent patterns contain crucial information about the physical and biological processes that occur in the systems in which they form \citep{Sole_2006, Meron_2018,Zhao2019}. 

\begin{figure}[!h]
    \centering
        \includegraphics[width=0.8\textwidth]{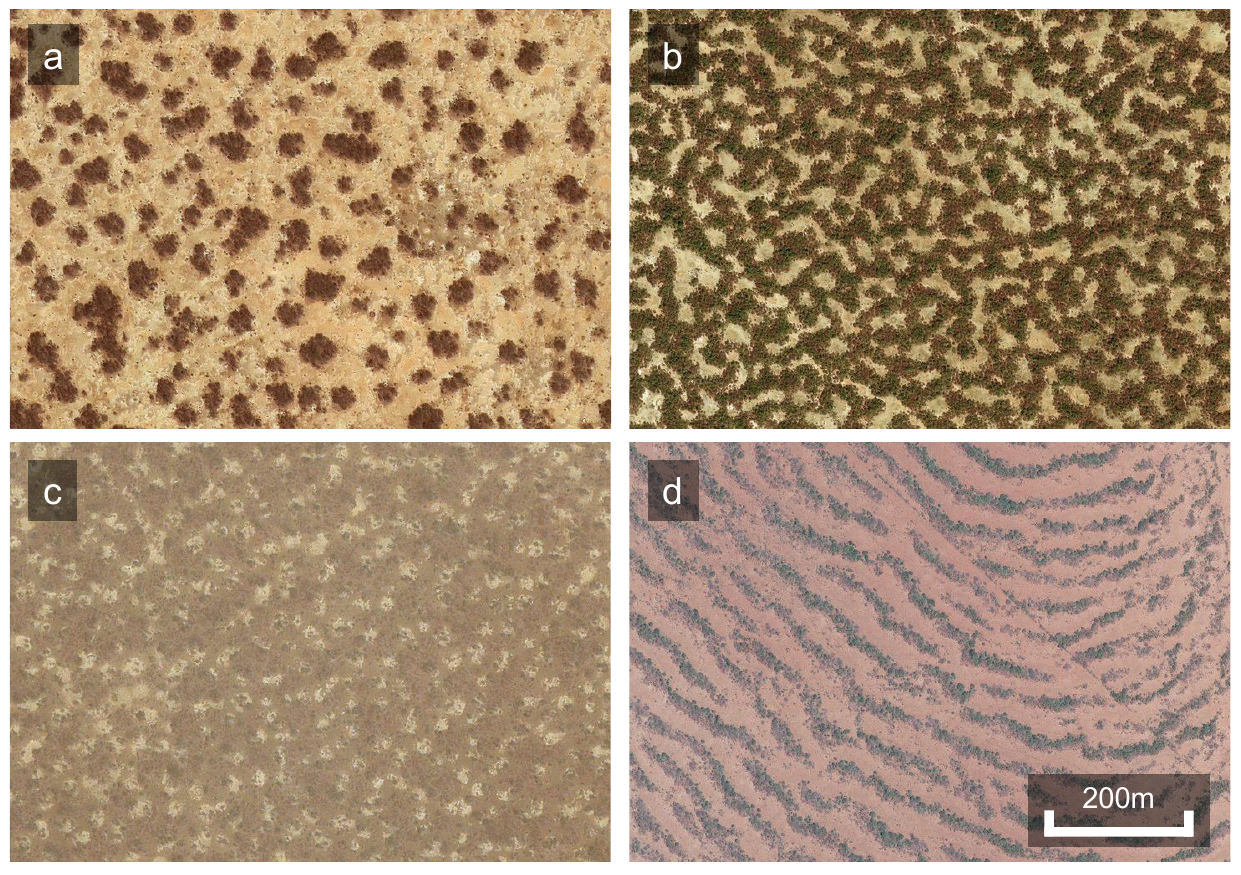}
        \caption{Aerial images of self-organized vegetation patterns. In all panels, vegetated regions are darker and bare-soil regions lighter. a) Spot pattern in Sudan (11$^\circ$34'55.2'' N; 27$^\circ$54'47.52''E). b) Labyrinthine pattern in Mali (12$^\circ$27'50''N; 3$^\circ$18'30''E). c) Gap pattern in Niger (11$^\circ$1'12''N; 28$^\circ$10'48''E). d) Band pattern in Sudan  (11$^\circ$3'0''N; 28$^\circ$17'24''E). Microsoft product screen shot(s) reprinted with permission from Microsoft Corporation. Image Copyrights $\copyright$2021 Maxar.}
        \label{fig:patterns}
\end{figure}

A well-known example of ecological self-organization is vegetation pattern formation in water-limited ecosystems \citep{Deblauwe_2008, Rietkerk2008}. Flat landscapes can show vegetation spots regularly distributed on a matrix of bare soil, soil-vegetation labyrinths, and gaps of bare soil regularly interspersed throughout a homogeneous layer of vegetation (Fig.\,\ref{fig:patterns}). Importantly, water availability strongly influences the specific shape of the pattern. In agreement with model predictions \citep{von_Hardenberg_2001, Meron_2004}, a Fourier-based analysis of satellite imagery covering extensive areas of Sudan revealed that more humid regions are dominated by gapped patterns, whereas spotted patterns dominate in more arid conditions \citep{Deblauwe2011}. However, imagery time series are in general not long enough to observe whether vegetation cover in a specific region undergoes these transitions between patterns when aridity increases over time (but see \citep{Bastiaansen2018}).  

Following the spotted pattern, if precipitation continues to decrease, models predict that patterned ecosystems undergo a transition to a desert state. This observed correlation between pattern shape and water availability suggests that the spotted pattern could serve as a reliable and easy-to-identify early-warning indicator of this ecosystem shift \citep{Lefever_1997,Rietkerk2004, Scheffer2009, Dakos2011, Dakos2015}. This potential application of vegetation patterns as early-warning indicators of ecosystem shift has reinforced the motivation to develop several models aiming to explain both pattern formation and their dependence on environmental parameters \citep{von_Hardenberg_2001,Meron_2004,Rietkerk_2002,Borgogno_2009,Martinez_Garcia_2014,Gowda2014}. Although there have been some attempts to test model predictions with satellite imagery and remote-sensed indices \citep{Bastiaansen2018,Weissmann2017,Veldhuis2021}, model-based theoretical studies remain the dominant approach to study this hypothesized transition.

All existing models successfully reproduce the sequence of gapped, labyrinthine, and spotted patterns found in satellite imagery (Fig.\,\ref{fig:1}a) \citep{Lejeune1999,Rietkerk_2002,Meron_2004,Martinez_Garcia_2014,Kealy2012}. However,
further analyses have found that each of this model predict a different desertification transition following the spotted pattern. For example, \citet{Rietkerk_2002} and \citet{von_Hardenberg_2001} predict that ecosystems undergo abrupt desertification, including a hysteresis loop, following the spotted pattern (bottom panel of Fig.\,\ref{fig:1}d). \citet{Siteur2014}, however, mathematically showed that patterns can be adaptive and change their periodicity in response to worsening environmental conditions. This study further showed that such pattern adaptability might help ecosystems avoid abrupt desertification processes, thus rendering the discontinuous transitions predicted by \citep{von_Hardenberg_2001} and \citep{Rietkerk_2002} continuous. \citet{Martinez_Garcia_2013} and \citet{Yizhaq2016} also predict that desertification could occur gradually with progressive loss of vegetation biomass (top panel of Fig.\,\ref{fig:1}d). Finally, \citet{Bastiaansen2020} suggested that patterned ecosystems collapse abruptly when environmental changes are fast and smoothly when the environmental conditions change slowly. Using alternative modeling approaches, other studies have supported the idea that whether an ecosystem will collapse gradually or abruptly is system-dependent and determined by the intensity of stochasticity \citep{Weissmann2017}, vegetation and soil type \citep{Kefi2007}, colonization rates \citep{Corrado2015}, and intensity of external stresses, such as grazing \citep{Kefi2007b}. This system dependence complicates the assessment and prediction of ecosystem health. Because drylands cover $\sim 40\%$ of Earth's land surface and are home to $\sim 35\%$ of the world population \citep{Mortimore2009}, determining whether these ecosystems will respond abruptly or gradually to aridification is critical both from ecosystem-management and socio-economic points of view.

Active lines of theoretical research aiming to address this question have focused on understanding how different components of the ecosystem may interact with each other to determine its response to aridification \citep{Bonachela2015,Yizhaq2016}, as well as on designing synthetic feedbacks, in the form of artificial microbiomes, that could prevent ecosystems collapses or make such transitions smoother \citep{Martin2015,Conde2020, Vidiella2020}. The question has also attracted considerable attention from empirical researchers \citep{Maestre2016}, and recent evidence suggests that certain structural and functional ecosystem attributes respond abruptly to aridity \citep{Berdugo2020}. Despite current efforts, whether desertification is more likely to occur gradually or abruptly remains largely unknown \citep{Rietkerk2021}.

Here, we discuss how self-organized vegetation patterns may help understand desertification processes. To this end, we review existing theoretical models and outline potential empirical approaches that will help test these models and ultimately elucidate how ecosystems respond to aridification. In section \ref{sec:rationale}, we discuss the ecological rationale behind existing models for vegetation self-organization. We review such models in section \ref{sec:models}, and  summarize their opposing predictions about the ecosystem collapse in section \ref{sec:transitions}. In section \ref{sec:experiments}, we describe possible manipulative experiments and empirical measures that can help when selecting among the previously scrutinized models. Finally, in section \ref{sec:conclusions}, we discuss different research lines that build on current knowledge and discuss how to apply lessons learned from studying self-organized vegetation patterns to other self-organizing systems. 

\section{Ecological rationale behind current models for vegetation spatial self-organization}\label{sec:rationale}

\begin{figure}[H]
    \centering
        \includegraphics[width=0.9\textwidth]{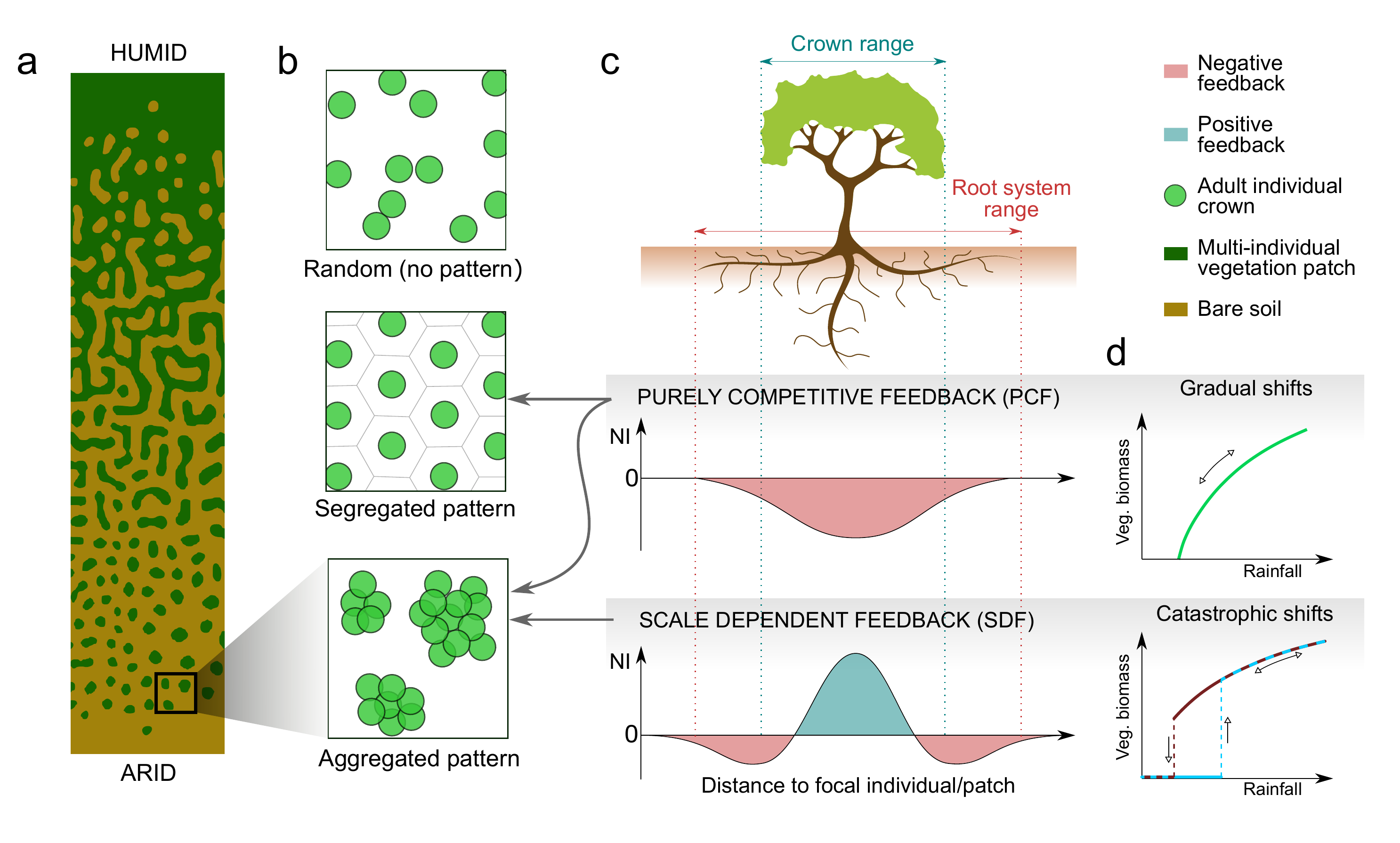}
        \caption{Conceptual summary of existing theories for vegetation self-organization, their emergent patterns and the type of desertification processes they predict. a) Graphical representation of observed spatial patterns across a gradient of rainfall, with more humid ecosystems showing gaps and more arid systems featuring spots. b) Examples of random, segregated, and aggregated patterns as defined in the main text. Aggregated patterns may result both from PCF and SDF, whereas segregated patterns emerge from PCF. c) Types of feedbacks invoked to explain the emergence of self-organized vegetation patterns and d) the different desertification transitions they predict. NI in the axes labels accounts for net interaction. \label{fig:1}}
\end{figure}

Depending on the number of individuals that form each patch, it is possible to classify non-random vegetation spatial patterns into two families: segregated and aggregated patterns (Fig.\,\ref{fig:1}b). We define a segregated pattern as a spatial distribution of vegetation in which plant biomass arranges in a usually hexagonal lattice of patches, with each individual representing one patch. Segregated patterns are very common in drylands and are expected to emerge exclusively from ecological interference or competition \citep{Tilman1997,Pringle_2017}. On the other hand, we define an aggregated pattern as a spatial distribution of vegetation in which plants form spatially-distant patches with several individuals in each patch. Aggregated patterns, which are the focus of our review, are ecologically more intriguing than segregated patterns because they might result from a richer set of mechanisms acting at different spatial scales \citep{Koppel_2008,Rietkerk2008, Lee2021}. This diversity of potential pattern causes is important because, at the ecosystem level, the ecological implications of the pattern strongly depend on the nature of the underlying mechanisms (Fig.\,\ref{fig:1}c,\,d). Moreover, direct evidence of which feedback type drives aggregated patterns of vegetation in drylands remains elusive.

Existing theories to explain the emergence of self-organized aggregated patterns are based on the biophysical effects that the plant canopy and the root system exert on the microclimate underneath and on the soil conditions, respectively \citep{Cabal2020} (Fig.\,\ref{fig:1}c). These theories ultimately connect these biophysical effects to the net biotic interaction between plants, i.e. the effect that plants have on each other’s survival, reproduction, and growth rate, and how it changes with inter-individual distance. To guide our discussion, we classify these theories in two broad categories, depending on whether they assume that the net biotic interaction between plants changes from positive to negative with increasing inter-individual distance (scale-dependent feedback; SDF) or remains negative at all spatial scales and only weakens with inter-individual distance (purely competitive feedback; PCF). Importantly, these two categories can be seen as the extremes of a continuum of models characterized by the ratio between the intensity of short-range positive and negative interactions. Next, we briefly review the mechanisms that have been suggested to underpin models within each of these two categories and the type of patterns that might emerge from them.

\textit{Scale-dependent feedbacks}. Biotic facilitation is a very common interaction in semiarid and arid ecosystems \citep{Holmgren2010}. The SDFs invoked to explain vegetation self-organization are caused by the facilitative effects of plants nearby their stems coupled with negative effects at longer distances. Several ecological processes have been suggested to support these SDFs. One is the positive effects of shade, which can overcome competition for light and the effects of root competition for water, and lead to under-canopy facilitation \citep{Valladares2016}. In this context, focal plants have an overall facilitative effect in the area of most intense shade at the center of the crown. This effect progressively loses intensity with distance to the center of the crown and ultimately vanishes, leaving just simple below-ground competition in areas farther from the plant. A complementary rationale is that plants modify soil crust, structure, and porosity, and therefore enhance soil water infiltration \citep{Eldridge2000,Ludwig2004}. Enhanced water infiltration has a direct positive effect close to the plant because it increases soil water content but, as a by-product, it has negative consequences farther away from the plant's insertion point because, by increasing local infiltration, plants also reduce the amount of water that can infiltrate in distant bare soil locations \citep{Montana1992,Bromley1997}. The spatial heterogeneity in water availability due to plant-enhanced infiltration is higher in sloped landscapes where down-slope water runoff leads to the formation of banded vegetation patterns \citep{Deblauwe2012,Valentin1999}, but it is also significant in flat landscapes and might cause the of emergence gaps, labyrinths, and spots of vegetation \citep{HilleRisLambers2001,Gilad2004,Okayasu2001}. (Fig.\,\ref{fig:1}a).

\textit{Purely competitive feedbacks}. Competition for resources is a ubiquitous interaction mechanism that affects the relation between any two plants when the distance between them is short enough. Above ground, plants compete for light through their canopies; below ground, they compete for several soil resources, including water and nitrogen, through their roots \citep{Craine2013}. If only competitive mechanisms occur, we should expect plants to have a negative effect on any other plant within their interaction range (top panel in Fig.\,\ref{fig:1}c) and the intensity of this effect to peak at intermediate distances between vegetation patches \citep{Koppel_2006}. Because finite-range competition is the only interaction required by PCF models to explain vegetation self-organization, PCF is the most parsimonious feedback type that generates vegetation patterns. 

\section{Models for vegetation self-organization}\label{sec:models}

Mathematical frameworks for self-organized vegetation patterns are grouped into two main categories: individual-based models (IBM) and partial differential equations models (PDEM). IBMs describe each plant as a discrete entity whose attributes change in time following a stochastic updating rule \citep{DeAngelis_2016,Railsback2019}. PDEMs describe vegetation biomass and water concentration as continuous fields that change in space and time following a system of deterministic partial differential equations \citep{Meron2015}. IBMs are the most convenient approach to study segregated patterns, where single individuals are easy to identify and central to the formation of vegetation patches \citep{Bolker1999,Iwasa2010, Wiegand2013,Plank2015}. Conversely, PDEMs are a better approximation to aggregated patterns because they focus on a continuous measure of vegetation abundance and describe the dynamics of patches that can spread or shrink without any upper or lower limit on their size. Natural multi-individual patches can change in size and shape depending on the interaction among the individual plants within them, whereas the size of single-plant patches is subject to stronger constraints (they usually grow, not shrink, and their maximum size is bounded by plant physiology). Therefore, PDEMs represent a simplification of the biological reality that is more accurate in aggregated than in segregated patterns. Because here we only consider aggregated patterns, we will focus our review of the mathematical literature on PDEMs. Within PDEMs, we first discuss reaction-diffusion formalisms, based on a system of coupled equations describing the dynamics of vegetation and water. Second, we dicuss kernel-based models that only describe the dynamics of vegetation biomass density and encapsulate all interactions between vegetation patches in an effective kernel function. This kernel function can model only competitive interactions (kernel-based PCF models; center panel of Fig.\,\ref{fig:1}c) or short-range facilitation and long-range inhibition (kernel-based SDF models; bottom panel of Fig.\,\ref{fig:1}c).

\subsection{Reaction-diffusion SDF models}\label{sec:sdfwater}

\begin{figure}[!h]
    \centering
        \includegraphics[width=0.8\textwidth]{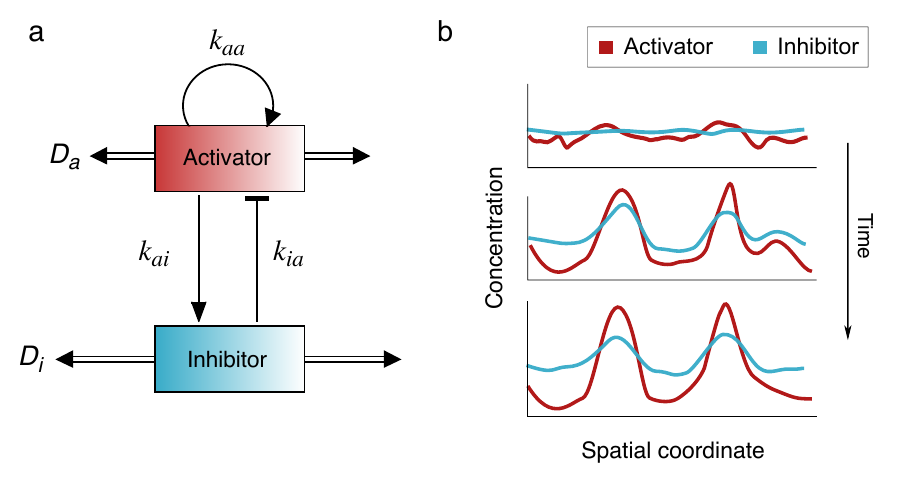}
        \caption{a) Schematic of the Turing activation-inhibition principle. The activator, with diffusion coefficient $D_a$, produces the inhibitor at rate $k_{ai}$ as well as more of itself at rate $k_{aa}$ through an autocatalytic reaction. The inhibitor degrades the activator at rate $k_{ia}$ and diffuses at rate $D_i > D_a$. b) Schematic of the pattern-forming process in a one-dimensional system.}
        \label{fig:turing}
\end{figure}

In 1952, Turing showed that differences in the diffusion coefficients of two reacting chemicals can lead to the formation of stable spatial heterogeneities in their concentration \citep{Turing1952}. In Turing's original model, one of the chemicals acts as an activator and produces both the second chemical and more of itself via an autocatalytic reaction. The second substance inhibits the production of the activator and therefore balances its concentration (Fig.~\ref{fig:turing}a). Spatial heterogeneities in the concentrations can form if the inhibitor diffuses much faster than the activator, so that it inhibits the production of the activator at a long range and confines the concentration of the activator locally (Fig.~\ref{fig:turing}b). This activation-inhibition principle thus relies on a SDF to produce patterns: positive feedbacks (autocatalysis) dominate on short scales and negative, inhibitory feedbacks dominate on larger scales. In the context of vegetation pattern formation, plant biomass acts as the self-replicating activator. Water is a limiting resource and thus water scarcity is an inhibitor of vegetation growth \citep{Rietkerk2008,Meron2012}.

\subsubsection{Two-equation water-vegetation dynamics: the generalized Klausmeier model}

Initially formulated to describe the formation of stripes of vegetation in sloping landscapes (Fig.\,\ref{fig:patterns}d) \citep{Klausmeier_1999}, subsequent studies have generalized the Klausmeier model to flat surfaces \citep{Kealy2011,Bastiaansen2018,Eigentler2020a}. Mathematically, the generalized version of the Klausmeier model is given by the following equations:
\begin{eqnarray}
\frac{\partial w({\bm r},t)}{\partial t}&=&R - l\,w({\bm r},t) - a\,g\left(w\right)f\left(v\right)v({\bm r},t) + D_w\mathrm{\nabla}^2 w({\bm r},t),	 \label{eq:gen-klaus-w}\\
\frac{\partial v({\bm r},t)}{\partial t}&=&a\,q\,g\left(w\right)f\left(v\right)v({\bm r}, t) - m\,v({\bm r}, t) + D_v\mathrm{\nabla}^2 v({\bm r},t), \label{eq:gen-klaus-v}
\end{eqnarray}
where $w({\bm r}, t)$ and $v({\bm r}, t)$ represent water concentration and density of vegetation biomass at location ${\bm r}$ and time $t$,  respectively. In Eq.~(\ref{eq:gen-klaus-w}), water is continuously supplied at a precipitation rate $R$, and its concentration decreases due to physical losses such as evaporation, occurring at rate $l$ (second term), and local uptake by plants (third term). In the latter, $a$ is the plant absorption rate, $g(w)$ describes the dependence of vegetation growth on water availability, and $f(v)$ is an increasing function of vegetation density that represents the positive effect that the presence of plants has on water infiltration. Finally, water diffuses with a diffusion coefficient $D_w$.  Similarly, Eq.\,(\ref{eq:gen-klaus-v}) accounts for vegetation growth due to water uptake (first term), plant mortality at rate $m$ (second term), and plant dispersal (last term). In the plant growth term, the parameter $q$ represents the yield of plant biomass per unit of consumed water; although the original model assumed for mathematical simplicity linear absorption rate and plant response to water (i.e., $g(w) = w({\bm r},t)$ and $f(v) = v({\bm r},t)$),  other biologically-plausible choices can account for processes such as saturation in plant growth due to intraspecific competition \citep{Eigentler2020}. 

The generalized Klausmeier model with linear absorption rate and plant responses to water has three spatially-uniform equilibria obtained from the fixed points of Eqs.\,(\ref{eq:gen-klaus-w})-(\ref{eq:gen-klaus-v}): an unvegetated state $(v^*=0, w^*=R/l)$, stable for any value of the rainfall parameter $R$; and two states in which vegetation and water coexist at different non-zero values. Only the vegetated state with higher vegetation biomass is stable against non-spatial perturbations, and only for a certain range of values of $R$. The latter suffices to guarantee bistability, that is, the presence of alternative stable states (vegetated vs unvegetated), and hysteresis. For spatial perturbations, however, the stable vegetated state becomes unstable within a range of $R$, and the system develops spatial patterns.

\subsubsection{Three-equation water-vegetation dynamics: the Rietkerk model}\label{sec:Rietkerk}

The Rietkerk model extends the generalized Klausmeier model by splitting Eq.\,(\ref{eq:gen-klaus-w}) into two equations (one for surface water, and another one for soil water) and including a term that represents water infiltration (see also \citep{Okayasu2001}). Moreover, the functions that represent water uptake and infiltration are nonlinear, which introduces additional feedbacks between vegetation, soil water, and surface water. The model equations are as follows:
\begin{eqnarray}
 \frac{\partial u(\mathbf{r},t)}{\partial t} &=& R - \alpha\, \frac{v(\mathbf{r},t) + k_2\,w_0}{k_2+v(\mathbf{r},t)}\,u(\mathbf{r},t) + D_u \nabla^2 u(\mathbf{r},t) \label{eq:surw}\\
 \frac{\partial w(\mathbf{r},t)}{\partial t} &=& \alpha\, \frac{v(\mathbf{r},t) + k_2\, w_0}{k_2 + v(\mathbf{r},t)}\,u(\mathbf{r},t) - g_{m}\, \frac{v(\mathbf{r},t)\, w(\mathbf{r},t)}{k_1 + w(\mathbf{r},t)} - \delta_w \, w(\mathbf{r},t) + D_w \nabla^2 w(\mathbf{r},t)  \label{eq:soilw} \\
  \frac{\partial v(\mathbf{r},t)}{\partial t} &=& c\, g_{m} \, \frac{v(\mathbf{r},t)\,w(\mathbf{r},t)}{k_1 + w(\mathbf{r},t)} - \delta_v \, v(\mathbf{r},t) + D_v \nabla^2 v(\mathbf{r},t) \label{eq:veg} 
\end{eqnarray}
where $u(\mathbf{r},t)$,  $w(\mathbf{r},t)$,  and $v(\mathbf{r},t)$ are the density of surface water, soil water, and vegetation, respectively. In Eq.\,(\ref{eq:surw}), $R$ is the mean annual rainfall (mm day$^{-1}$), providing a constant supply of water to the system; the second term accounts for infiltration; and the diffusion term accounts for the lateral circulation of water on the surface. In Eq.\,(\ref{eq:soilw}), the first term represents the saturating infiltration of surface water into the soil, which is enhanced by the presence of plants; the second term represents water uptake; the third one accounts for physical losses of soil water, such as evaporation; and the diffusion term describes the lateral circulation of underground water.  Finally, the first term in Eq.\,(\ref{eq:veg}) represents vegetation growth due to the uptake of soil water, which is a function that saturates for high water concentrations; the second term accounts for biomass loss at constant rate due to natural death or external hazards; and the diffusion term accounts for plant dispersal. The meaning of each parameter in the equations, together with the values used in \citet{Rietkerk_2002} for their numerical analysis, are provided in Table \ref{table_1}.

In the spatially uniform case, this model allows for two different steady states: a vegetated state in which vegetation, soil water, and surface water coexist at non-zero values; and an unvegetated (i.e., desert) state in which only soil water and surface water are non-zero. Considering the parameterization in Table \ref{table_1}, the stability of each of these states switches at $R=1$. For $R<1$, only the unvegetated equilibrium is stable against non-spatial perturbations, whereas for $R > 1$ the vegetated equilibrium becomes stable and the desert state, unstable. When allowing for spatial perturbations, numerical simulations using the parameterization in Table \ref{table_1} show the existence of spatial patterns within the interval $0.7 \lesssim R \lesssim 1.3$, which is in agreement with analytical approximations \citep{Gowda2016}. Within this range of mean annual rainfall, the patterns sequentially transition from gaps to labyrinths to spots with increasing aridity. For $R\approx 0.7$, the system transitions abruptly from the spotted pattern to the desert state.

\begin{table}[ht]
\centering 
\begin{tabular}{|c|c|c|} 
\hline 
Parameter & Symbol & Value \\ [0.5ex] 
\hline 
 $c$					& Water-biomass conversion factor & $10$ (g mm$^{-1}$ m$^{-2}$) \\
 $\alpha$			& Maximum infiltration rate &  $0.2$ (day$^{-1}$) \\
 $g_{m}$			& Maximum uptake rate & $0.05$ (mm g$^{-1}$ m$^{-2}$ day$^{-1}$) \\
 $w_0$				& Water infiltration in the absence of plants &  $0.2$ (-) \\
 $k_1$ 				& Water uptake half-saturation constant & $5$ (mm) \\
 $k_2$				& Saturation constant of water infiltration & $5$ (g m$^{-2}$)  \\
 $\delta_w$     & Soil water loss rate &  $0.2$ (day$^{-1}$) \\
 $\delta_v$ 		& Plant mortality & $0.25$  (day$^{-1}$) \\
 $D_w$ 			& Soil water lateral diffusion & $0.1$ (m$^2$ day$^{-1}$) \\
 $D_v$ 				& Vegetation dispersal & $0.1$ (m$^2$ day$^{-1}$) \\
 $D_u$ 			& Surface water lateral diffusion & $100$ (m$^2$ day$^{-1}$)\\
\hline 
\end{tabular}
\caption{Typical parameterization of the Rietkerk model \citep{Rietkerk_2002}.} 
\label{table_1} 
\end{table}

In its original formulation, the Rietkerk model assumes constant rainfall, homogeneous soil properties, and only local and short-range processes. Therefore, all the parameters are constant in space and time, and patterns emerge from SDFs between vegetation biomass and water availability alone. This simplification is, however, not valid for most ecosystems. Arid and semi-arid regions feature seasonal variability in rainfall \citep{Salem1989} and depending on the functional dependence between water uptake and soil moisture, stochastic rainfall might increase the amount of vegetation biomass in the ecosystem compared to a constant rainfall scenario \citep{Kletter2009}. Moreover, the properties of the soil often change in space.  A widespread cause of this heterogeneity is soil-dwelling macrofauna, such as ants, earthworms, and termites \citep{Pringle_2017}. Heterogeneity in substrate properties induced by soil-dwelling macrofauna, and modeled by space-dependent parameters, might interact with SDFs between water and vegetation and introduce new characteristic spatial scales in the pattern \citep{Bonachela2015}. Finally, researchers have also extended the Rietkerk model to account for long-range, nonlocal processes. For example, a nonlocal mechanism in the interaction between vegetation biomass and soil water of Eqs.\,(\ref{eq:soilw})-(\ref{eq:veg}) can model the water conduction of lateral roots towards the plant canopy \citep{Gilad2004}. As explained in the next section, although the model in \citep{Gilad2004} accounts for nonlocal processes, it is conceptually very different from kernel-based models.

\subsection{Kernel-based SDF models}\label{sec:kernelsdf}

Kernel-based models are those in which all the feedbacks that control the interaction between plants are encapsulated in a single nonlocal net interaction between plants. The nonlocality in the net plant interaction accounts for the fact that individual (or patches of) plants can interact with each other within a finite neighborhood. Therefore, the vegetation dynamics at any point of the space is coupled to the density of vegetation at locations within the interaction range. Because all feedbacks are collapsed into a net interaction between plants, kernel-based models do not describe the dynamics of any type of water and use a single partial integro-differential equation for the spatiotemporal dynamics of the vegetation. The kernel is often defined as the addition of two Gaussian functions with different widths, with the wider function taking negative values to account for the longer range of competitive interactions \citep{DOdorico2006} (center plot in Fig.\,\ref{fig:1}c).

\subsubsection{Models with additive nonlocal interactions}

In the simplest kernel-based SDF models, the spatial coupling is introduced linearly in the equation for the local dynamics \citep{DOdorico2006},
\begin{equation}\label{eq:kernel1}
\frac{\partial v({\bm r},t)}{\partial t}=h\left(v\right)+\int d{\bm r'} G\left({\bm r'};{\bm r} \right)\left[v\left({\bm r'},t\right)-v_0\right].
\end{equation}
The first term describes the local dynamics of the vegetation, i.e., temporal changes in vegetation density at a location ${\bm r}$ due to processes in which neighboring vegetation does not play any role. The integral term describes any spatial coupling, i.e., changes in vegetation density at ${\bm r}$ due to vegetation density at neighbor locations ${\bm r}$'. Assuming spatial isotropy, the kernel function $G({\bm r}, {\bm r'})$ decays radially with the distance from the focal location, $\rvert {\bm r'} - {\bm r}\rvert$, so $G\left({\bm r'},{\bm r}\right)=G(\rvert {\bm r'} - {\bm r}\rvert)$. Therefore, two main contributions govern the dynamics of vegetation density: first, if the spatial coupling is neglected, the vegetation density increases or decreases locally depending on the sign of $h(v)$ until reaching a uniform steady state $v_0$, solution of $h(v_0)=0$; second, the spatial coupling enhances or diminishes vegetation growth depending on the sign of the kernel function (i.e., whether the spatial interactions affect growth positively or negatively) and the difference between the local vegetation density and the uniform steady state $v_0$. 

Assuming kernels that are positive close to the focal location and negative far from it (modeling a SDF), local perturbations in the vegetation density around $v_0$ are locally enhanced if they are larger than $v_0$ and attenuated otherwise. As a result, the integral term destabilizes the homogeneous state when perturbed, and spatial patterns arise in the system. Long-range growth-inhibition interactions, together with nonlinear terms in the local-growth function $h(v)$, avoid the unbounded growth of perturbations and stabilize the pattern. However, although this mechanism imposes an upper bound to vegetation density, nothing prevents $v$ from taking unrealistic, negative values. To avoid this issue, the model must include an artificial bound at $v = 0$ such that vegetation density is reset to zero whenever it becomes negative.

\subsubsection{Models with multiplicative nonlocal interactions}

A less artificial way to ensure that vegetation density remain always positive is to modulate the spatial coupling with nonlinear terms. For example, the pioneering model developed by \citet{Lefever_1997} consists of a single integro-differential equation describing the spatio temporal dynamics of vegetation biomass, $v$,
\begin{equation}\label{eq:lefever}
\frac{\partial v({\bm r}, t)}{\partial t} = \beta \,  \left[ \omega_1 * (v \,(1 + \Omega\, v) \right]({\bm r},t)  \left[ 1 - \frac{\left( \omega_2 * v \right)({\bm r}, t)}{K}\right]  - \eta \left( \omega_3 * v \right)({\bm r},t) 
\end{equation}
\noindent Eq.\,(\ref{eq:lefever}) is a modified logistic equation with seed production rate $\beta$ and an additional facilitation term modulated by the parameter $\Omega$. 
$\eta$ is the rate at which vegetation biomass is lost due to spontaneous death and external hazards such as grazing, fires, or anthropogenic factors (last term). Each of the terms includes a spatial convolution that encodes long-range spatial interactions via a weighted average of vegetation density within a neighborhood of the focal location,
\begin{equation}\label{eq:conv}
\left( \omega_i * f(v) \right)({\bm r},t) = \int d{\bm r}' \omega_i({\bm r}-{\bm r}'; \ell_i) \, f_i[v({\bm r}', t)] \ \ \ \ \ \ \ \ \ \ \ \ \ \ \ \ \ \ \ \ \ \ \ \ \ \ \ \ \ \ i=1,2,3
\end{equation} 
In the convolutions defined by Eq.\,(\ref{eq:conv}), the model assumes spatial isotropy. Each $\omega_i$ is a weight function with a characteristic spatial scale $\ell_i$ that defines the size of the neighborhood contributing to the focal process. For instance, $\omega_1({\bm r}-{\bm r}'; \ell_1)$ defines $\ell_1$ as linear the size of the neighborhood that contributes to the growth of vegetation biomass at ${\bm r}$. Similarly, $\ell_2$ defines the scale over which plants inhibit the growth of their neighbors, and $\ell_3$ the scale over which vegetation density influences the spontaneous death rate of vegetation at the focal location. These scales, however, are rarely parameterized using field measurements of root extent and crown size (but see \citep{Couteron2014,Tlidi2018}). Instead, they are set to meet the conditions of a scale-dependent feedback with long-range growth inhibition and short-range activation. In \citet{Lefever_1997}, the authors assume $\ell_2>\ell_1$, and the model includes a SDF with short-range facilitation and long-range competition.

Expanding upon this work, other studies have introduced non-linear spatial couplings via spatial convolutions multiplying only some of the terms of the model equation \citep{Fernandez_Oto_2014,Escaff_2015,Berrios2020}, and others have expanded the integral terms and studied the formation of localized structures of vegetation \citep{Escaff_2015,Lejeune2002,Parra2020}.

\subsection{Kernel-based PCF models.}\label{sec:compmodels}
In previous sections, we invoked the existence of SDFs in the interactions among plants to explain the emergence of self-organized spatial patterns of vegetation. Both theoretical and empirical studies, however, have highlighted the importance of long-range negative feedbacks on pattern formation, suggesting that short-range positive feedbacks might be secondary actors that sharpen the boundaries of clusters rather than being key for the instabilities that lead to the patterns \citep{Koppel_2006,Rietkerk2008,Koppel_2008}. Following these arguments, \citet{Martinez_Garcia_2013}, \citep{Martinez_Garcia_2014} proposed a family of purely competitive models with the goal of identifying the smallest set of mechanisms needed for self-organized vegetation patterns to form.
 
The simplest PCF models consider additive nonlocal interactions \citep{Martinez_Garcia_2014}. Alternatively, nonlocal interactions can be represented through nonlinear functions modulating either the growth or the death terms. In both cases, the models develop the full sequence of patterns (gaps, labyrinths, and spots). The model proposed by \citet{Martinez_Garcia_2013} introduces competition through the growth term:
\begin{equation}\label{eq:nonl-comp}
\frac{\partial v({\bm r},t)}{\partial t}=P_{\mbox{\tiny{E}}} \left(\widetilde{v}, \delta\right)\beta\, v({\bm r},t)\left(1-\frac{v({\bm r}, t)}{K}\right)-\eta \, v({\bm r},t),
\end{equation}
where $\beta$ and $K$ are the seed production rate and the local carrying capacity, respectively. $P_E$ is the probability that seeds overcome competition and establish as new vegetation biomass. $\delta$ is the competition-strength parameter, and $\widetilde{v}\left({\bm r},t\right)$ is the average density of vegetation around the focal position ${\bm r}$:
\begin{equation}\label{eq:nonlocal-ker}
\widetilde{v}\left({\bm r}, t\right)=\int d{\bm r}' \omega \left(\rvert {\bm r'} - {\bm r}\rvert \right)\, v\left({\bm r}',t\right).
\end{equation}
where the kernel function $\omega$ is a weight function and thus plays the same role and has the same properties described for the $\omega_i$ functions in section \ref{sec:kernelsdf}. The model further assumes that vegetation losses occur at constant rate $\eta$ and vegetation grows through a three-step sequence of seed production, local dispersal, and establishment \citep{Calabrese2010}, represented by the three factors that contribute to the first term in Eq.\,(\ref{eq:nonl-comp}). First, plants produce seeds at a constant rate $\beta$, which leads to the a growth term $\beta v({\bm r},t)$.  Second, seeds disperse locally and compete for space which defines a local carrying capacity $K$. Third, plants compete for resources with other plants, which is modeled using the plant establishment probability, $P_{\mbox{\tiny{E}}}$.  Because the only long-range interaction in the model is root-mediated interference, and competition for resources is more intense in more crowded environments,  $P_E$ is a monotonically decreasing function of the nonlocal vegetation density $\tilde{v}({\bm r},t)$ defined in Eq.\,(\ref{eq:nonlocal-ker}).  Moreover, $P_{\mbox{\tiny{E}}}$ also depends on the competition-strength parameter, $\delta$, representing resource limitation. In the limit $\delta = 0$, resources are abundant, competition is weak, and $P_{\mbox{\tiny{E}}} =1$. Conversely, in the limit $\delta\rightarrow\infty$, resources are very scarce, competition is very strong, and $P_{\mbox{\tiny{E}}}\rightarrow 0$.

\begin{figure}[!h]
    \centering
        \includegraphics[width=0.8\textwidth]{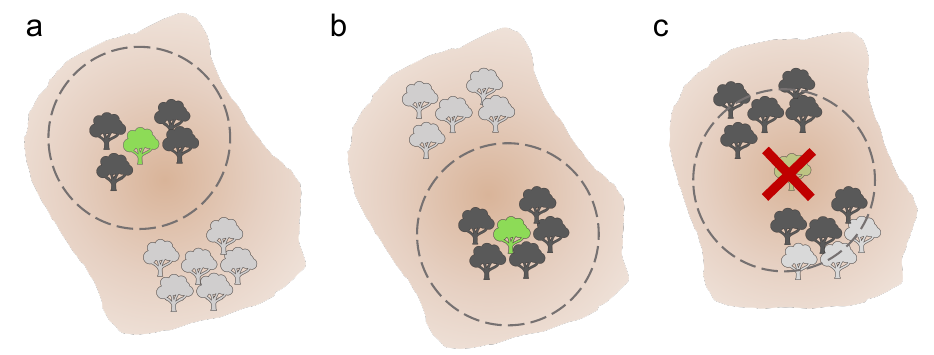}
        \caption{In PCF models, patchy distributions of vegetation in which the distance between patches is between one and two times the range of the nonlocal interactions are stable. Individuals within each patch only compete with the individuals in that patch (a,b), whereas individuals in between patches compete with individuals from both patches (c). Color code: green trees are focal individuals, and dashed circles limit the range of interaction of the focal individual. Dark grey is used for individuals that interact with the focal one, whereas light gray indicates individuals that are out of the range of interaction of the focal individual.}
        \label{fig:exclusion}
\end{figure}

In PCF models, spatial patterns form solely due to long-range competition. If the characteristic range of competition is comparable to the inter-patch distance, individuals attempting to establish in between patches compete with vegetation from more than one adjacent patch, whereas individuals within a patch only interact with plants in that same patch. As a result, competition is more intense in the regions between patches than inside each patch, which stabilizes an aggregated pattern of vegetation (Fig.~\ref{fig:exclusion}) whose shape (gaps, labyrinths or spots) will depend on the model parameterization.  This same mechanism has been suggested to drive the formation of clusters of competing species in niche space \citep{Scheffer_2006,Pigolotti2007,Hernandez-Garcia2009,Fort2009,Leimar2013} and the aggregation of individuals in models of interacting particles with density-dependent reproduction rates \citep{Hernandez-Garcia2004}.

\section{Self-organized patterns as indicators of ecological transitions}\label{sec:transitions}

Models using different forms for the net biotic interaction between neighbor patches (SDF vs PCF) have succeeded at reproducing qualitatively the spatial patterns of vegetation observed in water-limited ecosystems \citep{Deblauwe2011}. All these different models also predict that a spotted pattern precedes a transition to an unvegetated state, suggesting that clumped vegetation patterns could be an early-warning indicator of desertification transitions \citep{Scheffer2009, Dakos2011}. Models assuming different underlying mechanisms for the formation of these patterns, however, result in different desertification processes.

\begin{figure}[H]
    \centering
        \includegraphics[width=0.8\textwidth]{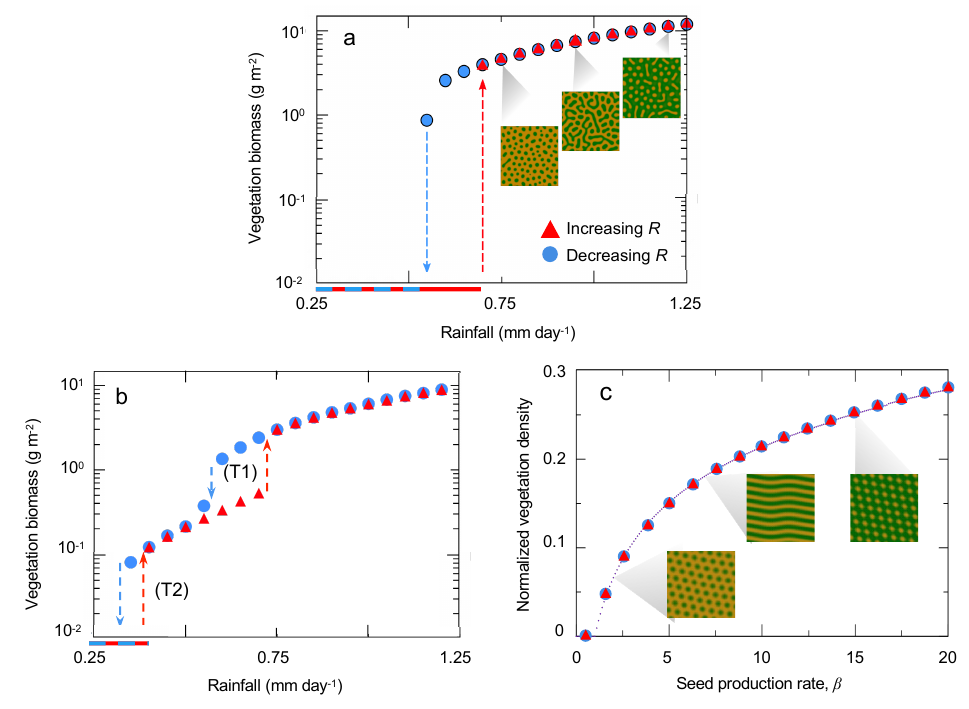}
        \caption{Although different models for vegetation pattern formation may recover the same sequence of gapped-labyrinthine-spotted patterns from different mechanism, the type of desertification transition that follows the spotted pattern strongly depends on the model ingredients. a) Abrupt desertification as predicted by the Rietkerk model \citep{Rietkerk_2002}. b) Two-step desertification process as predicted in \citet{Bonachela2015} when soil-dwelling insects are added to the Rietkerk model. c) Progressive desertification as predicted by the PCF model introduced in \citet{Martinez_Garcia_2013}. For each panel, numerical simulations were conducted using the same setup described in the original publications.}
        \label{fig:transition}
\end{figure}

The Rietkerk model from section \ref{sec:Rietkerk}, for example, predicts that self-organized ecosystems eventually collapse following an abrupt transition that includes a hysteresis loop (Fig.~\ref{fig:transition}a). Abrupt transitions such as this one are typical of bistable systems in which the stationary state depends on the environmental and initial conditions. Bistability is a generalized feature of models for vegetation pattern formation, sometimes occurring also in transitions between patterned states \citep{von_Hardenberg_2001}. It also denotes the existence of thresholds in the system that trigger sudden, abrupt responses in its dynamics. These thresholds are often created by positive feedbacks or quorum-regulated behaviors, as is the case in populations subject to strong Allee effects \citep{Courchamp1999}.  In the Rietkerk model, as rainfall decreases, the spatial distribution of vegetation moves through the gapped-labyrinthine-spotted sequence of patterns (Fig.~\ref{fig:transition}a). However, the system responds abruptly when the rainfall crosses a threshold value, and all vegetation dies. Setting up the simulations as indicated by \citet{Rietkerk_2002} and using the parameterization of Table ~\ref{table_1},  this threshold is located at $R\approx 0.55$ mm day$^{-1}$. Once the system reaches this unvegetated state, increasing water availability does not allow vegetation recovery until $R\approx 0.70$ mm day$^{-1}$, which results in a hysteresis loop and a region of bistability ($R\in [0.55, 0.70]$ in Fig.~\ref{fig:transition}a). Bistability and hysteresis loops make abrupt, sudden transitions like this one extremely hard to revert. Hence, anticipating such abrupt transitions is crucial from a conservation and ecosystem-management point of view \citep{Scheffer2009, Dakos2011}. 

Extended formulations of the Rietkerk model have suggested that the interaction between vegetation and other biotic components of the ecosystem may change the transition to the unvegetated state. Specifically, soil-dwelling termites, in establishing their nests (mounds), engineer the chemical and physical properties of the soil in a way that turns the abrupt desertification into a two-step process (Fig.~\ref{fig:transition}b) \citep{Bonachela2015}. At a certain precipitation level ($R\approx 0.75$ mm day$^{-1}$ using the parameterization in Table \ref{table_1} and the same initial condition used for the original Rietkerk model), vegetation dies in most of the landscape (T1 in Fig. \ref{fig:transition}b) but persists on the mounds due to improved properties for plant growth created by the termites. On-mound vegetation survives even if precipitation continues to decline, and is finally lost at a rainfall threshold $R\approx 0.35$ mm day$^{-1}$ (T2 in Fig. \ref{fig:transition}b). As a consequence of the two-step transition, the ecosystem collapse is easier to prevent, because a bare soil matrix with vegetation only on mounds serves as an early-warning signal of desertification, and it is easier to revert because termite-induced heterogeneity breaks the large hysteresis loop of the original model into two smaller ones (compare the hysteresis loops in Figs.~\ref{fig:transition}a and \ref{fig:transition}b).

Alternative stable states can also occur among patterns. The generalized Klausmeier model exhibits a range of stable patterned states for a given rainfall value, and thus pattern multistability \citep{Siteur2014}. Therefore, observed patterns only contain partial information about the ecosystem state since they will be very strongly determined by the system history. Likewise, this model predicts that a pattern with a specific periodicity is stable for a wide range of rainfall values, which means that it could persist if environmental conditions worsen as long as they remain within the range in which the pattern is stable. As a general rule, the generalized Klausmeier model predicts that patterns with higher wavenumbers are stable at higher rainfall values, and lower wavenumbers become stable as rainfall decreases. Therefore, this model predicts that patterned ecosystems can respond to decreasing water availability by adjusting the pattern wavenumber, which can eventually allow the system to evade abrupt desertification processes \cite{Siteur2014}. This possibility is more likely if environmental conditions change slowly \citep{Bastiaansen2020}. More recent work has hypothesized that a similar multistability could appear in the Rietkerk model, challenging the importance of spatial patterning as an early-warning indicator of abrupt desertification processes and regime shifts \citep{Rietkerk2021}. 

Finally, the PCF model discussed in section \ref{sec:compmodels} predicts a smooth desertification in which vegetation biomass decreases continuously in response to decreasing seed production rate (a proxy for worsening environmental conditions) \citep{Martinez_Garcia_2013b}. According to this model, the spotted pattern would persist as precipitation declines, with vegetation biomass decreasing in the patch until it eventually disappears (Fig.~\ref{fig:transition}c). Different from the catastrophic shift described for the Rietkerk model, smooth transitions such as the one depicted by this model do not show bistability and do not feature hysteresis loops. This difference has important socio-ecological implications, because it enables easier and more affordable management strategies to restore the ecosystem after the collapse \citep{Martin2015}. Moreover, continuous transitions are also more predictable because the density of vegetation is univocally determined by the control parameter (seed production rate, $\beta$ in Fig.\,\ref{fig:transition}c) and does not depend on the system history. Importantly, this specific model assumes that vegetation dispersal is local, and hence Eq.\,(\ref{eq:nonl-comp}) does not include a diffusion term. Further bifurcation analysis should be performed to determine whether the spotted pattern bifurcates from the unvegetated state following a supercritical transition due to the absence of a positive feedback or the lack of diffusion.

In the three scenarios discussed above, self-organized vegetation patterns appear as an inexpensive and reliable early indicator of ecological transitions \citep{Scheffer2009, Dakos2011}. We have shown that widespread spotted patterns can form in models accounting for very different mechanisms (Fig.\,\ref{fig:transition}). Therefore, the different predictions that models make about the transition highlights the need for tailored models that not only  reproduce the observed patterns but do so through mechanisms relevant to the focal system.  Because ecosystems are highly complex, it is very likely that spotted patterns observed in different regions emerge from very different mechanisms (or combinations of them) and thus anticipate very different transitions \citep{Kefi2007b,Kefi2007, Corrado2015, Weissmann2017}. Therefore, a reliable use of spotted patterns as early warning indicators of ecosystem collapse requires (i) mechanistic models that are parameterized and validated by empirical observations of both mechanisms and patterns; (ii) quantitative analyses of field observations involving as many variables as possible; and (iii) manipulative experiments. 

\section{Testing models for vegetation self-organization in the field} \label{sec:experiments}

As evidenced through this review, the study of self-organized vegetation patterns has been mostly theoretical and empirical evidence of the self-organization hypotheses explaining the formation of vegetation spatial patterns are much less widespread. In this section, we explore possible empirical approaches to understanding the mechanisms responsible for the formation and persistence of self-organized aggregated patterns and propose a two-step protocol to this end. 

At least three possible approaches exist to testing existing self-organization hypotheses for pattern formation in natural systems. First, the observation and measurement of the spatial structure of patterns from aerial and satellite photography (the observational approach). Within this observational approach, we can distinguish between a single objective optimization approach that only aims to match pattern shapes, and a multiobjective optimization approach in which models must simultaneously be consistent with multiple output variables (pattern shape, vegetation density, water infiltration rates...) when properly parameterized. Second, the assessment of the net interactions among plants or plant patches as a function of the distance between them (the net-interaction approach). Third, the investigation of the specific mechanisms underpinning that interaction (the selective mechanistic approach). The observational approach, mainly on its single-objective optimization variant, has been relatively common, and is the one that has motivated the development of most existing models \citep{Borgogno_2009,Wu2000}. Nevertheless, without a more detailed understanding of the focal systems, one cannot discard whether agreement between model predictions and natural observations is coincidental. The mechanistic approach has been relatively common. Some studies have investigated selectively the mechanisms potentially leading to pattern formation in tiger bush banded vegetation in Niger \citep{Valentin1999b}, plant tussock and cushions in the Andean Altiplano \citep{Couteron2014}, vegetation rings in Israel \citep{Sheffer2011,Yizhaq2019}, or fairy circles in Namibia \citep{Ravi2017} and Australia \citep{Getzin2021}. To our knowledge, the net-interaction approach is conspicuous by being absent, and researchers have scarcely measured directly the net interaction among plants or vegetation patches and its variation with the inter-plant distance in patterned ecosystems. 

Because we are discussing here aggregated patterns, a first step in such an approach is to confirm that vegetation patches are formed by several aggregated individuals. In the case of segregated patterns, we recommend the use of point pattern analysis under the hypothesis of the dominance of competition forces \citep{Franklin2010}. Following this preliminary test, we propose a two-step protocol to conduct future field research on the emergence and maintenance of vegetation spatial patterns. 

\begin{figure}[!h]
    \centering
        \includegraphics[width=0.7\textwidth]{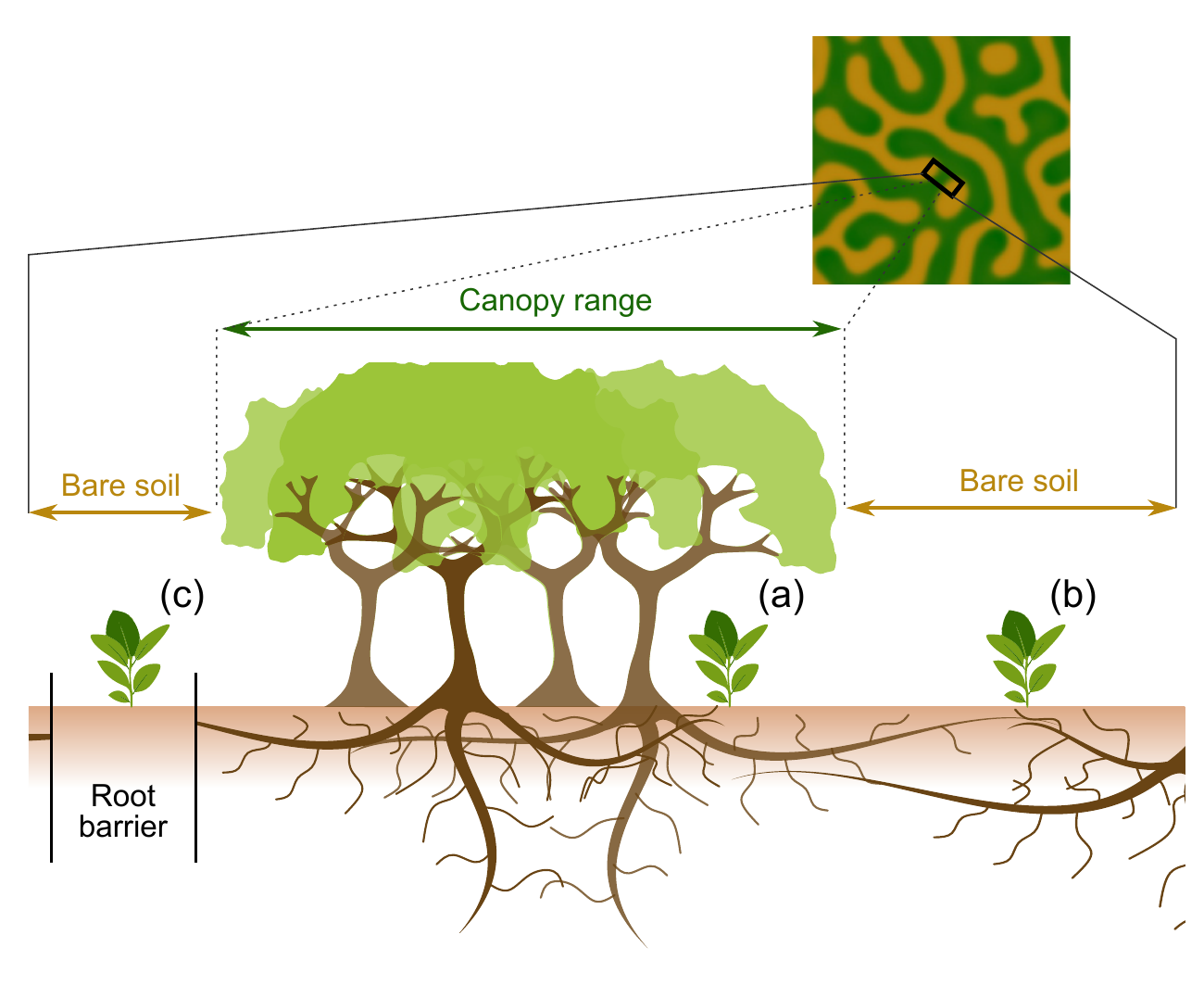}
        \caption{Schematic representation of a simple experimental setup to test in the field whether the mechanism of spatial patterning is a purely competitive feedback (PCF) or a classic scale-dependent feedback (SDF). Plant (a) is an experimental plant growing under-canopy,  (b) is growing in bare soil, and (c) is a control plant growing in artificial conditions, free from the biotic interaction using root barriers in bare soil areas of the same environment. }
        \label{fig:experiment}
\end{figure}

\textit{First step: phenomenological investigation of the net interaction among plants within the vegetated patch and in bare soil.} This first step is needed because a myriad of alternative mechanisms can explain the formation of spatial patterns, some of them well aligned with the idea of vegetation self-organization and others dependent on external biological or geological factors. For example, in the case of fairy circles researchers have explored the role of higher evaporation \citep{Vlieghe2019} and increased termite activity \citep{Juergens2013} within the circles; spatial heterogeneity in hydrological processes, such as increased infiltration in the circles of bare soil \citep{Ravi2017} or increased water runoff in the circles and in the matrix \citep{Getzin2021}; and the geological emanation of toxic gases \citep{Naude2011} and the presence of allelochemical substances \citep{Meyer2015} in the circles. By using a phenomenological approach as a first step, researchers can discard many of these potential mechanisms, directing their efforts towards more specific hypotheses in a second step. For this first step, a simple experimental setup, based on mainstream methods to measure plant biotic interaction would reveal whether PCF, SDF, or none of them are responsible for the emergent pattern \citep{Armas2004}. Our proposed experiment would compare a fitness proxy (e.g., growth, survival) of experimental plants planted in the system under study, where we observe a regular vegetation pattern and we assume that the vegetation vegetation dynamics has reached a stationary state for the location-specific environmental conditions. Each experimental block would consist of a plant growing under-canopy (Fig.\,\ref{fig:experiment}a), a plant growing in bare soil (i.e., between two vegetation patches) (Fig.\,\ref{fig:experiment}b), and a control plant growing in the same ecosystem but artificially isolated from the interaction with pattern-forming individuals (Fig.\,\ref{fig:experiment}c). To isolate control plants from canopy interaction they need to be planted in bare soil areas. To isolate them from below-ground competition, one can excavate narrow, deep trenches in which a root barrier can be inserted \citep{Morgenroth2008}. To isolate them from the competition for runoff water with the vegetation patches, root barriers should protrude from the soil a few centimeters, preventing precipitation water to leave the area and runoff water to enter. Comparing the fitness proxy of the control plant with that of plants growing in vegetation patches or bare soil reveals the sign and strength of the net biotic interaction. By replicating this experimental block, we can statistically determine whether the pattern formation results from a SDF, a PCF, or whether it involves a different process. The SDF hypothesis would be validated if a predominantly positive net interaction is observed under the canopy, and a negative interaction is observed in bare soils. Conversely, the PCF would be validated if we observe a negative net interaction in bare soils and under canopy (see Table \ref{table_2}). Any other outcome in the spatial distribution of the sign of the net interaction between plants would suggest that other mechanisms are at play, which could include the action of different ecosystem components, such as soil-dwelling macrofauna \citep{Tarnita2017}, or abiotic factors, such as micro-topology.

\begin{table}[ht]
\centering 
\begin{tabular}{|c|c|c|} 
\hline 
Under canopy vs control				& Bare soil vs control & Outcome \\
\hline 
 $0 / -$			& $ - \ \ -$ &  Purely competitive feedback \\
 $+$			& $ - $ & Scale-dependent feedback \\
\hline 
\end{tabular}
\caption{Testing the PCF versus SDF hypotheses in the experimental setup introduced in Fig.\,\ref{fig:experiment}. Double signs indicate stronger intensity. Indexes to calculate the sign of the net interaction can be taken from \citet{Armas2004}.} 
\label{table_2} 
\end{table}

\textit{Second step: direct measurement of the biophysical processes responsible for the pattern.} After confirming the PCF, SDF, finding an alternative spatial distribution of plant interactions, or rejecting the self-organizing hypothesis, a second experimental step would test specific biophysical mechanisms responsible for the measured interaction and driving the spatial pattern. For example, PCF models hypothesize that spatial patterns are driven by long-range below-ground competition for a limiting resource through the formation of exclusion regions. As discussed in section \ref{sec:compmodels}, these exclusion regions are territories between patches of vegetation in which the intensity of competition is higher than within the patch \citep{Koppel_2006}, possibly because they present a higher density of roots (Fig.\, \ref{fig:exclusion}) \citep{Martinez_Garcia_2013, Martinez_Garcia_2014}. To test for the existence of exclusion regions and confirm whether below-ground competition is driving the spatial pattern, researchers can measure the changes in root density using coring devices \citep{Cabal2021} across soil transects going from the center of a vegetated patch to the center of a bare soil patch. Field tests and manipulative experiments to confirm that SDFs are responsible for vegetation patterns are not easy to perform but researchers can do a handful of analyses. For example, ecohydrological SDF models assume that water infiltration is significantly faster in vegetation patches than in bare soil areas \citep{Rietkerk_2002}. Many empirical researchers have tested this assumption in patterned vegetation using mini disk infiltrometers to quantify unsaturated hydraulic conductivity, dual head infiltrometers to measure saturated hydraulic conductivity, or buried moisture sensors connected to data loggers to record volumetric soil moisture content \citep{Yizhaq2019,Ravi2017,Getzin2021,Cramer2013}. The measures should show higher infiltration rates and moisture under vegetated patches than in bare soil. Note, however, that infiltration rates might be very hard to measure due to small-scale soil heterogeneities. Ecohydrological models make other assumptions that are less often considered in the field. For instance, they assume that the lateral transport of water is several orders of magnitude larger than vegetation diffusion (i.e., patch growth speed), which might be true or not depending on soil properties. To test these assumptions, field researchers need to measure the intensity of the water runoff and compare it to a measure of the lateral growth rate of vegetation patches. Water runoff is very challenging to measure directly, but estimates can be calculated using the infiltration rates obtained with infiltrometers \citep{Cook1946}. The lateral growth rate of vegetation patches can be estimated based on drone or satellite images repeated over time \citep{Trichon2018}. Combining measures of both water runoff and expansion rates of vegetation patches, one can estimate approximated values for the relative ratio of the two metrics.

\section{Conclusions and future lines of research}\label{sec:conseq}\label{sec:conclusions}

As our ability to obtain and analyze large, high-resolution images of the Earth's surface increases, more examples of self-organized vegetation patterns are found in water-limited ecosystems. Here, we have reviewed different modeling approaches employed to understand the mechanistic causes and the predicted consequences of those patterns. We have discussed how different models, relying on different mechanisms, can successfully reproduce the patterns observed in natural systems despite the fact that each of these models predicts very different ecosystem-level consequences of the emergent pattern. This discrepancy limits the utility of vegetation patterns as applied ecological tools. To solve this issue we propose a new approach to studying vegetation pattern formation in water-stressed systems. This new approach should abandon the development of phenomenological models that are validated qualitatively via the visual comparison of simulated and observed (macroscopic) patterns and pursue a more mechanistic and system-specific description of vegetation self-organization. This new approach must necessarily focus on isolating the system-specific, key feedbacks for vegetation self-organization. To achieve this goal, we identify two main directions for future research.

In the first direction, we propose to extend the current model validation based on comparing simulated and observed pattern to other ecological variables that can be predicted by existing models. Recent developments in remotely sensed imagery have enabled the measurement of an ecosystem's state indicators, which will allow researchers to compare observed and simulated patterns quantitatively and to extend this comparison to other model outputs such as soil moisture or water infiltration rates. This more comprehensive comparison between simulations and data would allow researchers to conduct a model selection analysis based on a multi-objective optimization process and thus classify existing models from more to less realistic depending on whether (and how many) features of the focal ecosystem the model manages to reproduce in the correct environmental conditions. For example, models could be classified depending on whether, after proper parameterization, they can predict ecosystem responses such as transitions between pattern types at the correct aridity thresholds. To elaborate this model classification, the use of Fourier analysis for identifying regularity in natural patterns, geostatistics for quantifying spatial correlations, and time series analysis for tracking changes in the ecosystem properties through time will be essential. 

In the second direction, biologically-grounded studies should aim to combine system-specific models with empirical measures of vegetation-mediated feedbacks.  Experimental measures of the (microscopic) processes and feedbacks central to most models of vegetation pattern formation are hard to obtain, leading to arbitrary (free) parameter values and response functions. For example, very few models incorporate empirically-validated values of water diffusion and plant dispersal rates, despite the crucial role of these parameters in the emergence of patterns. Instead, these models fine-tune such values to obtain patterns similar in, for example, their wavelength, to the natural pattern. Similarly, we are only beginning to understand how plants rearrange their root system in the presence of competing individuals \citep{Cabal2020b,Cabal2021b}, and hence kernel-based models still do not incorporate realistic functional forms for the kernels. Instead, these models use phenomenological functions to test potential mechanisms for pattern formation by qualitatively comparing model output and target pattern, thus limiting the potential of the models to make quantitative predictions. To establish a dialogue between experiments and theory, these purely mechanistic models should develop from a microscopic description of the system \citep{DeAngelis_2016, Railsback2019}, which allows for a more realistic and accurate description of the plant-to-plant and plant-water interactions as well as for a better reconciliation between model parameters and system-specific empirical measures. Subsequently, existing tools from mathematics, statistical physics, and/or computer science can be used to reach a macroscopic PDEM that captures the key ingredients of the microscopic dynamics. Statistical physics, which was conceived to describe how observed macroscopic properties of physical systems emerge from the underlying microscopic processes, provides a compelling and well-developed framework to make such a micro-macro connection. 

Beyond water-limited ecosystems, SDFs have been invoked to explain spatial pattern formation in mussel beds \citep{Rietkerk2008}, freshwater and salt marshes \citep{VandeKoppel2005,VanWesenbeeck2008,Zhao2021}, and seagrasses \citep{VanderHeide2011,Ruiz-Reynes2017}. Conversely, nonlocal competition drives the emergence of aggregated patterns in freshwater marshes \citep{Koppel_2008} and in theoretical models of population dynamics \citep{Fuentes2003, Dornelas2019, Maruvka2006, DaCunha2011, Clerc2005, Maciel2020,Piva2021}. Understanding the conditions in which negative feedbacks dominate over positive feedbacks, finding the key features that distinguish the patterns caused by these different feedbacks, and contrasting their divergent ecological consequences constitutes an exciting venue for future research that has just started to develop \citep{Lee2021}.

\vspace{6pt} 
\subsection*{Acknowledgments}
We thank Robert M.  Pringle, Rub\'en Juanes, and Ignacio Rodr\'iguez-Iturbe for various discussions at different stages of the development of this work. This work is supported by FAPESP through grants ICTP-SAIFR 2016/01343-7, and Programa Jovens Pesquisadores em Centros Emergentes 2019/24433-0 and 2019/05523-8; ICTP through the Associates Programme; Simons Foundation through grant number 284558FY19 (RMG). The Princeton University May Fellowship in the department of Ecology and Evolutionary Biology (CC); the Spanish State Research Agency, through the Severo Ochoa and María de Maeztu Program for Centers and Units of Excellence in R\&D (MDM-2017-0711) funded by MCIN/AEI/10.13039/501100011033 (EHG \& CL). The NSF RoL:FELS:EAGER-1838331 (CET); Gordon and Betty Moore Foundation, grant \#7800 (CET \& JAB). NSF grant DMS-2052616; Simons Foundation (award \#82610) (JAB). This work was partially  funded by the Center of Advanced Systems Understanding (CASUS) which is financed by Germany’s Federal Ministry of Education and Research (BMBF) and by the Saxon Ministry for Science, Culture and Tourism (SMWK) with tax funds on the basis of the budget approved by the Saxon State Parliament (JMC).



\end{document}